# Implications for a spatially discrete transition amplitude in the twin-slit experiment

*W.M. Stuckey[1]*

**Abstract**

A discrete path integral formalism is used to obtain the transition amplitude between 'sources' (slits and detector) in the twin-slit experiment of quantum mechanics. This method explicates the normally tacit construct of dynamic entities with temporal duration. The resulting amplitude is compared to that of Schrödinger dynamics in order to relate 'source' dynamics and spatial separation. The implied metric embodies non-separability, in stark contrast to the metric of general relativity. Thus, this approach may have implications for quantum gravity.

[1] Department of Physics, Elizabethtown College, Elizabethtown, PA 17022-2298**,** stuckeym@etown.edu

# 1. INTRODUCTION

According to Feynman[1], the twin-slit experiment "has in it the heart of quantum mechanics. In reality, it contains the *only* mystery." Herein we address this "mystery" by taking to heart Pauli's admonition that[2] "in providing a systematic foundation for quantum mechanics, one should start more from the composition and separation of systems than has until now (with Dirac, e.g.) been the case." Our result resonates strongly with Smolin's belief[3] that what "we are all missing" in the search for quantum gravity "involves two things: the foundations of quantum mechanics and the nature of time."

We start in section 2 by introducing a discrete path integral formalism whence quantum mechanics (QM) follows in the temporally continuous and spatially discrete limits while quantum field theory (QFT) follows in the temporally and spatially continuous limits. Per this formalism we are able to explicate the manner in which relations (as opposed to the wave function) may be viewed as fundamental to relata (such as particles) as suggested by our previous work on the *Relational Blockworld* (RBW) interpretation[4] of QM. We believe that, contrary to convention, the fully spatiotemporally discrete starting point is more fundamental than its QM and QFT limits, since an explicit construct of the discrete action is required. For example, by relating our fully discrete Lagrangian to its temporally continuous QM counterpart, we expose the notion of trans-temporal identity[5] employed tacitly in the construct of an action (classical or quantum). We suggest the process of trans-temporal identification can be articulated via a self-consistency criterion relating dynamical/diachronic entites, space and time. This approach would constitute a *unification* of physics as opposed to a mere *discrete approximation* thereto, since we are proposing a source for the action, which is otherwise fundamental. This may shed light on the "nature of time" as necessary, per Smolin, for quantum gravity. We finish section 2 by obtaining the transition amplitude for two interacting QM 'sources', i.e., a source and detector in the parlance of QM. We use this result in section 3 to obtain the QM amplitude for the twin-slit experiment.

When we compare the spatially discrete transition amplitude to the amplitude obtained via Schrödinger dynamics, we find a relationship between spatial distance and 'source' dynamics quite unlike that of Einstein's equations of general relativity (GR). In particular, the implied metric isn't an "extreme embodiment of the separability principle[6]."

To wit, there are no waves or particles propagating from source to detector through fields or otherwise empty space during the exchange of momentum. Indeed, this notion of spatial distance is not defined between points of empty space, but only between interacting 'sources'. Thus, our rendition of the twin-slit experiment necessarily circumvents "a fundamental incompatibility between general relativity and quantum mechanics[7]," i.e., QM embodies non-separability via quantum entanglement while the GR metric and its underlying differentiable manifold embody pervasive spatiotemporal separability. In this sense, QM's "only mystery" may indeed be a foundational issue relevant to quantum gravity per Smolin's suggestion.

## 2. DISCRETE PATH INTEGRAL FORMALISM

To formalize the idea that spatial separation exists only between interacting trans-temporal objects we suggest a spatiotemporally discrete formalism underlying quantum physics with QM following in the spatially discrete, temporally continuous limit and QFT following in the limit of both spatial and temporal continuity (Figure 1). To motivate this approach, consider the QFT transition amplitude for sources interacting via a scalar field without scattering per Zee[8]

$$Z = \int D\varphi \exp\left[i\int d^4x \left[\frac{1}{2}(d\varphi)^2 - V(\varphi) + J(x)\varphi(x)\right]\right] \quad (1)$$

According to Zee, the QM counterpart then obtains in (0+1) dimensions. In the derivation of Eq. (1) *from* QM, the field φ is obtained in the continuum limit of a discrete set of oscillators $q_i$ distributed in a spatial lattice. Any *one* of these $q_i$ is supposed to replace φ in Eq. (1) in order that it reduce to QM. However, each $q_i$ is fixed in space so the notion that we're integrating over all possible paths *in space* (standard treatment) from a source to a detector when we compute Z (the propagator in QM) is not ontologically consistent with the fact that we integrate over all values of q but *not* over all values of the index 'i' in $q_i$. Thus, we suggest the method for obtaining QM is to associate the sources J(x) with elements in the experimental set up (all of which may be deemed "sources" *and* "detectors" in a relational reality) while maintaining a discrete collection $q_i(t)$. [We will see in section 3 how this differs from the "free-particle" propagator in the standard path integral approach to QM.]

**Figure 1**

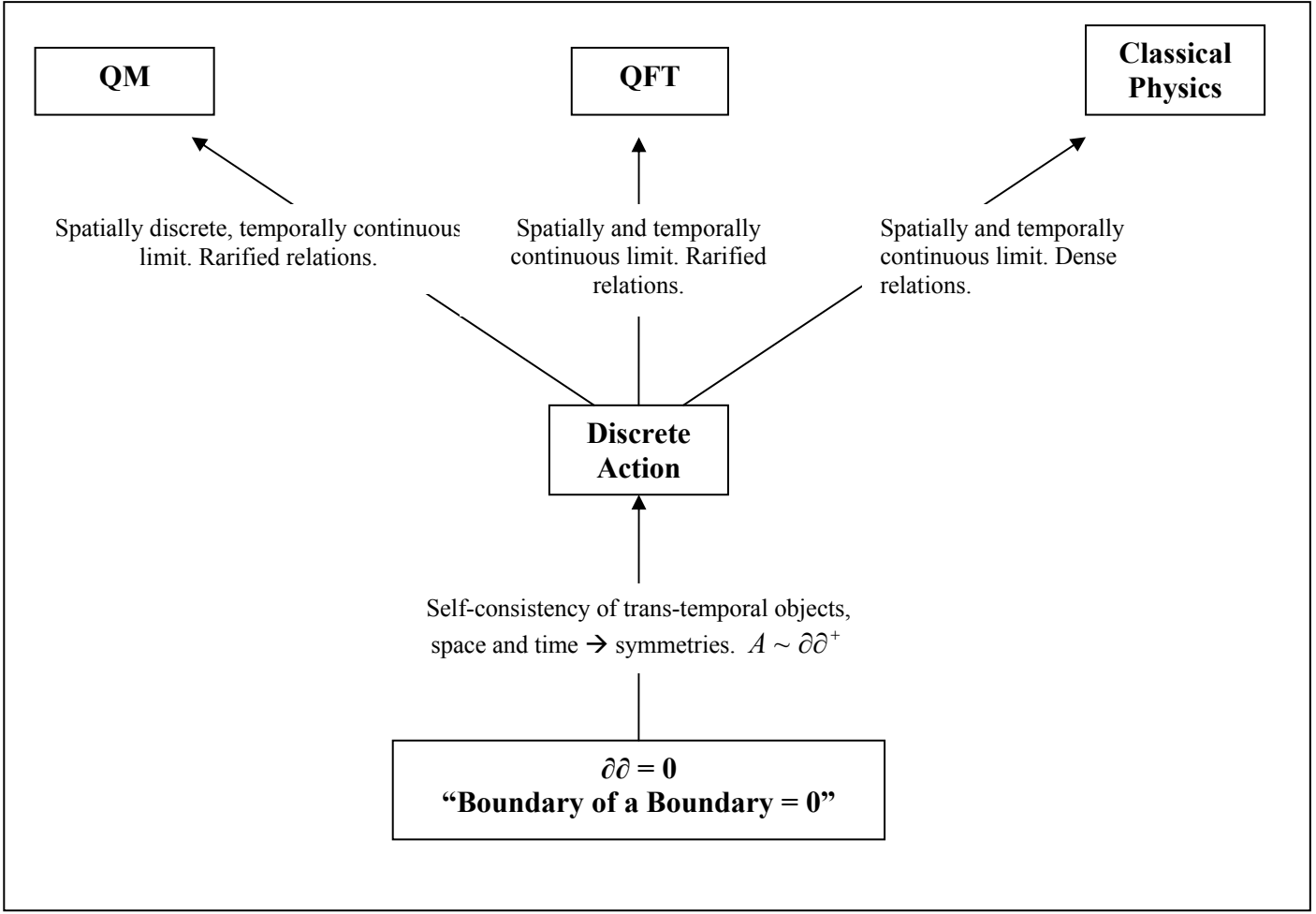

The spatiotemporally discrete counterpart to Eq. (1) is

$$Z = \int ... \int dQ_1...dQ_N \, \exp\left[\frac{i}{2} Q \cdot A \cdot Q + iJ \cdot Q\right] \qquad (2)$$

when V(φ) is quadratic, where $A_{ij}$ is the discrete matrix counterpart to the differential operator of Eq. (1) while $J_m$ and $Q_n$ are the discrete vector versions of $J(x)$ and $\varphi(x)$. The discrete action, $\frac{1}{2} Q \cdot A \cdot Q + J \cdot Q$, is considered a functional, which we may write as $\frac{1}{2}|\alpha\rangle\langle\beta| + \langle J|$, of $Q_n$, which we may write as $\langle Q|$ or $|Q\rangle$. Regions in $Q_n$ space where the action is stationary, i.e., invariant/symmetric, contribute most prominently to the transition

amplitude[a], which is now best viewed as a "symmetry amplitude" rather than a "particle propagator." Therefore, the functional is constructed so that what one means by trans-temporal objects, space and time, per $\langle J|$ and $|\alpha\rangle\langle\beta|$ respectively, are self-consistently defined and harbor the desired fundamental symmetries (Figure 1). This is of course similar to the modus operandi of theoretical particle physics, the difference being the discrete formalism allows for (requires) the explicit construct of trans-temporal objects in concert with the spacetime metric whereas the spatiotemporally continuous starting point of QFT harbors tacit assumptions/constraints[b].

The solution to Eq. (2) is

$$Z = \left(\frac{(2i\pi)^N}{\det(A)}\right)^{1/2} \exp\left[-\frac{i}{2} J \cdot A^{-1} \cdot J\right] \qquad (3)$$

Since $A_{ij}$ has an inverse, it has a non-zero determinant so it's composed of N linearly independent vectors in its N-dimensional, representational vector space. Thus, any vector in this space may be expanded in the set of vectors comprising $A_{ij}$. Specifically, the vector $J_m$, which will be used to represent 'sources' in the experimental set-up, can be expanded in the vectors of $A_{ij}$. In this sense it is clear how relations, represented by $A_{ij}$, can be fundamental to relata, represented by $J_m$. In the case of two coupled harmonic oscillators we have

$$V(q_1, q_2) = \sum_{a,b} \frac{1}{2} k_{ab} q_a q_b = \frac{1}{2} kq_1^2 + \frac{1}{2} kq_2^2 + k_{12} q_1 q_2$$

where $k_{11} = k_{22} = k$ and $k_{12} = k_{21}$, so our Lagrangian is

$$L = \frac{1}{2} m\dot{q}_1^2 + \frac{1}{2} m\dot{q}_2^2 - \frac{1}{2} kq_1^2 - \frac{1}{2} kq_2^2 - k_{12} q_1 q_2$$

and the spatially and temporally discrete version of $A_{ij}$ in Eq. (2) would be

$$A = - \begin{pmatrix} \frac{m}{\Delta t} + k\Delta t & \frac{-2m}{\Delta t} & \frac{m}{\Delta t} & 0 & \ldots & 0 & k_{12}\Delta t & 0 \\ 0 & \frac{m}{\Delta t} + k\Delta t & \frac{-2m}{\Delta t} & \frac{m}{\Delta t} & 0 & \ldots & 0 & k_{12}\Delta t \\ & & & \ddots & & & & \\ k_{12}\Delta t & 0 & \ldots & \frac{m}{\Delta t} + k\Delta t & \frac{-2m}{\Delta t} & \frac{m}{\Delta t} & 0 & \ldots \\ 0 & k_{12}\Delta t & 0 & \ldots & \frac{m}{\Delta t} + k\Delta t & \frac{-2m}{\Delta t} & \frac{m}{\Delta t} & 0 \\ & & & & & \ddots & & \end{pmatrix} \quad (4)$$

---

[a] Each possible experimental outcome of a given experiment requires its own spatiotemporally holistic description yielding its own transition amplitude. For the case of spatially discrete sources, $Z$ is the probability amplitude so it provides a frequency over the possible outcomes via the Born rule.

[b] That one must explicitly construct the trans-temporal objects, space and time of the discrete action suggests a level of formalism fundamental to the action. Toffoli[9] has proposed that a mathematical tautology resides at this more fundamental level, e.g., "the boundary of a boundary is zero" whence general relativity and electromagnetism[10]. In a forthcoming arXiv posting, we will use discrete graph theory to propose a self-consistency criterion which is also based on this tautology. Again, since we propose an explicit means for constructing the discrete action, we believe this discrete approach is a unification of physics rather than a mere discrete approximation thereto.

The process of temporal identification $Q_n \rightarrow q_i(t)$ may be encoded in the blocks along the diagonal of $A_{ij}$ whereby the spatial division between the $q_i(t)$ would then be encoded in the relevant off-diagonal (interaction) blocks:

$$A = \begin{pmatrix} \boxed{\begin{matrix} \ddots & & \\ & q_1(t) & \\ & & \ddots \end{matrix}} & \boxed{q_1(t) \Leftrightarrow q_2(t)} \\ \boxed{q_2(t) \Leftrightarrow q_1(t)} & \boxed{\begin{matrix} \ddots & & \\ & q_2(t) & \\ & & \ddots \end{matrix}} \end{pmatrix}$$

The discrete formulation illustrates nicely how QM tacitly assumes an *a priori* process of trans-temporal identification, $Q_n \rightarrow q_i(t)$. Indeed, there is no principle which dictates the construct of trans-temporal objects fundamental to the formalism of dynamics in general – these objects are "put in by hand." Thus, RBW suggests the need for a fundamental principle which would explicate the trans-temporal identity employed tacitly in QM, QFT and all dynamical theories. Since our starting point does not contain trans-temporal objects, space or time, we have to formalize counterparts to these concepts. Clearly, the process $Q_n \rightarrow q_i(t)$ is an organization of the set $Q_n$ on two levels—there is the split of the set into subsets, one for each 'source', and there is the ordering over each subset. The split represents space (true multiplicity from apparent identity), the ordering represents time (apparent identity from true multiplicity)[c] and the result is objecthood (via relations). Again, the three concepts are inextricably linked in our formalism, thus our suggestion that they be related via a self-consistency criterion (Figure 1).

In the limit of temporal continuity, Eq. (4) dictates we find the inverse of

$$-\begin{pmatrix} m\dfrac{d^2}{dt^2}+k & k_{12} \\ k_{12} & m\dfrac{d^2}{dt^2}+k \end{pmatrix}$$

to obtain Eq. (3) so that

$$-\frac{1}{2}Q \cdot A \cdot Q \to \int \left( \frac{m}{2} q_1 \ddot{q}_1 + \frac{1}{2}kq_1^2 + \frac{m}{2} q_2 \ddot{q}_2 + \frac{1}{2}kq_2^2 + k_{12}q_1 q_2 \right) dt$$

in our QM action. Solving

$$-\begin{pmatrix} m\dfrac{d^2}{dt^2}+k & k_{12} \\ k_{12} & m\dfrac{d^2}{dt^2}+k \end{pmatrix} D_{im}(t-t') = \begin{pmatrix} \delta(t-t') & 0 \\ 0 & \delta(t-t') \end{pmatrix}$$

for $D_{im}(t-t')$ we find

$$D_{im}(t-t') = -\begin{pmatrix} \int \dfrac{d\omega}{2\pi} A(\omega)e^{i\omega(t-t')} & \int \dfrac{d\omega}{2\pi} B(\omega)e^{i\omega(t-t')} \\ \int \dfrac{d\omega}{2\pi} B(\omega)e^{i\omega(t-t')} & \int \dfrac{d\omega}{2\pi} A(\omega)e^{i\omega(t-t')} \end{pmatrix}$$

with

$$A = \frac{\omega^2 m - k}{k_{12}^2 - (\omega^2 m - k)^2} \quad \text{and} \quad B = \frac{k_{12}}{k_{12}^2 - (\omega^2 m - k)^2}$$

The QM amplitude in this simple case is then given by

$$Z(J) \propto \exp\left[-\frac{i}{\hbar} \iint dt dt' J_1(t) D_{12} J_2(t') \right] = \exp\left[\frac{i}{\hbar} \iiint \frac{dt' dt d\omega}{2\pi} \frac{J_1(t) k_{12} e^{i\omega(t-t')} J_2(t')}{k_{12}^2 - (\omega^2 m - k)^2} \right]$$

---

[c] These definitions of space and time follow from a fundamental principle of standard set theory, *multiplicity iff discernibility* [11].

having restored $\hbar$, used $D_{12} = D_{21}$ and ignored the "self-interaction" terms $J_1 D_{11} J_1$ and $J_2 D_{22} J_2$. Fourier transforms give

$$Z(j) \propto \exp\left[\frac{i}{\hbar} \int \frac{d\omega}{2\pi} \frac{j_1(\omega)^* k_{12} j_2(\omega)}{\left(k_{12}^2 - (\omega^2 m - k)^2\right)}\right] \tag{5}$$

with $J_1(t)$ real.

## 3. TWIN-SLIT EXPERIMENT AND SEPARABILITY

If we now use this amplitude to analyze the twin-slit experiment, we can compare the result to that of Schrödinger dynamics and infer a spatial distance. There are four $J$'s which must be taken into account when computing the amplitude (figure 2), so we will use Eq. (5) to link $J_1$ with each of $J_2$ and $J_4$, and $J_3$ with each of $J_2$ and $J_4$, i.e., $J_1 \leftrightarrow J_2 \leftrightarrow J_3$ and $J_1 \leftrightarrow J_4 \leftrightarrow J_3$. In doing so, we ignore the contributions from other pairings, i.e., the exact solution would contain one integrand with $Q_n \rightarrow q_i(t)$, i = 1,2,3,4. Finally, we assume a monochromatic source of the form $j_1(\omega)^* = \Gamma_1 \delta(\omega - \omega_o)$ with $\Gamma_1$ a constant, so the amplitude between $J_1$ and $J_2$ is

$$Z(j) \propto \exp\left[\frac{i}{2\pi\hbar} \frac{\Gamma_1 k_{12} j_2(\omega_o)}{\left(k_{12}^2 - (\omega_o^2 m - k)^2\right)}\right]$$

whence we have for the amplitude between $J_1$ and $J_3$ via $J_2$ and $J_4$

$$\psi \propto \exp\left[\frac{i}{2\pi\hbar}(\Gamma_1 d_{12} j_2 + \Gamma_2 d_{23} j_3)\right] + \exp\left[\frac{i}{2\pi\hbar}(\Gamma_1 d_{14} j_4 + \Gamma_4 d_{43} j_3)\right] \tag{6}$$

where

$$d_{im} = \frac{k_{im}}{\left(k_{im}^2 - (\omega_o^2 m - k)^2\right)} \tag{7}$$

with $\psi$ the QM amplitude. [$Z$ corresponds to the QM propagator which yields the functional form of $\psi$ between spatially localized sources, as will be seen below.]

**Figure 2**

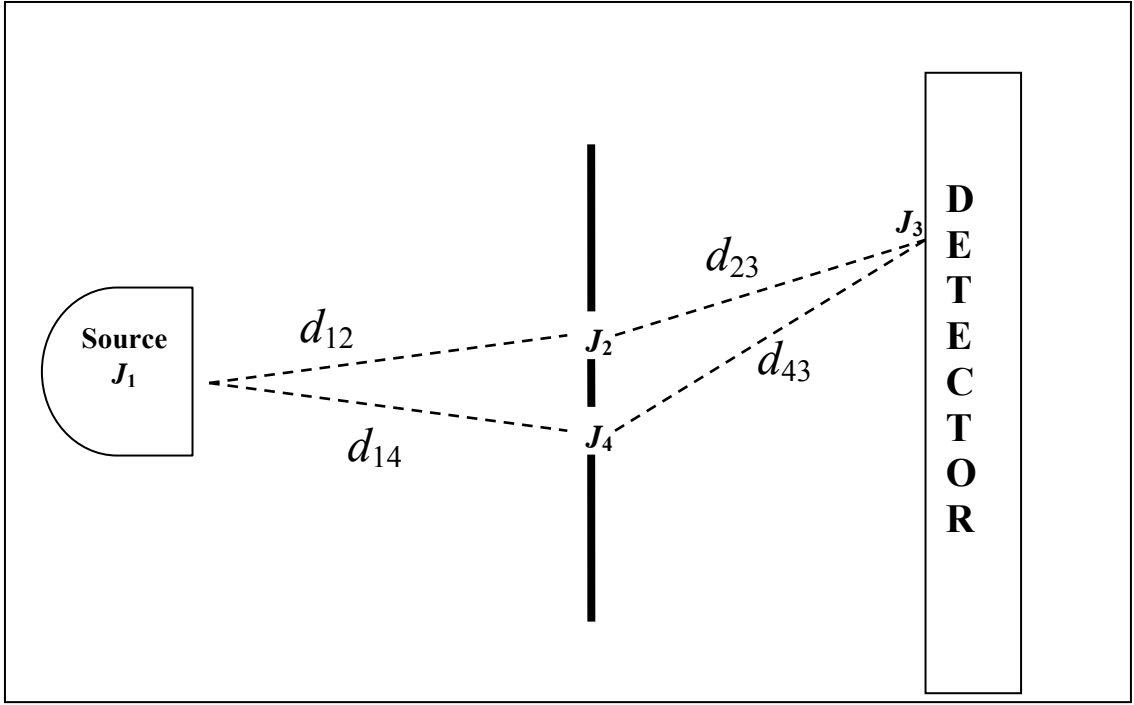

With the source equidistance from either slit (or, equivalently, with slits replaced by a pair of coherent laser-excited atoms[12]) the phase $\Gamma_1 d_{12}j_2$ equals the phase $\Gamma_1 d_{14}j_4$, so we have the familiar form

$$\psi \propto \exp\left[\frac{i}{2\pi\hbar}(\Gamma_2 d_{23} j_3)\right] + \exp\left[\frac{i}{2\pi\hbar}(\Gamma_4 d_{43} j_3)\right] \tag{8}$$

Now we need the corresponding result from Schrödinger dynamics. The free-particle propagator of Schrödinger dynamics is[13]

$$U(x_2,t;x_1,0) = \sqrt{\frac{m}{2\pi\hbar it}} \exp\left[\frac{im(x_2-x_1)^2}{2\hbar t}\right]$$

for a particle of mass $m$ moving from $x_1$ to $x_2$ in time $t$. This 'exchange' particle has no dynamic counterpart in the formalism used to obtain Eq. (8), but rather is associated with the oscillatory nature of the spatially discrete 'source' (see below). According to our view, this propagator is tacitly imbued "by hand" with notions of trans-temporal objects, space and time per its derivation via the free-particle Lagrangian. In short, the construct of this propagator bypasses explicit, self-consistent construct of trans-temporal objects, space and time thereby ignoring the self-consistency criterion fundamental to the action. The self-inconsistent, tacit assumption of a single particle with two worldlines (a "free-particle propagator" for each slit) is precisely what leads to the "mystery" of the twin-slit experiment. This is avoided in our formalism because $Z$ does not represent the propagation of a particle between 'sources', e.g., $q_i(t) \neq x(t)$ as explained *supra*. Formally, the inconsistent, tacit assumption is reflected in $-\frac{1}{2} Q \cdot A \cdot Q \rightarrow \int \left(\frac{m}{2} \dot{x}^2\right) dt$ where ontologically *m* (which is *not* the same *m* that appears in our oscillator potential) is the mass of the 'exchange' particle (i.e., purported dynamical/diachronic entity moving between 'sources' – again, the ontic status of this entity is responsible for the "mystery") and $x(t)$ (which, again, is *not* equal to $q_i(t)$) is obtained by *assuming* a particular spatial metric (this assumption *per se* is not responsible for the "mystery"). Its success in producing an acceptable amplitude when integrating over all paths $x(t)$ in space ('wrong' techniques can produce 'right' answers), serves to deepen the "mystery" because the formalism, which requires interference between different spatial paths, is not consistent with its antecedent ontological assumption, i.e., single particle causing a single click. There is no such self-inconsistency in our approach, because $Z$ is not a "particle propagator" but a 'mathematical machine' which measures the degree of symmetry contained in the spatiotemporally holistic configuration of trans-temporal objects, space and time represented by $A$ and $J$, as explained *supra*. Thus, this QM "mystery" results from an attempt to tell a dynamical story in an adynamical situation. Continuing, we have

$$\psi(x_2, t) = \int U(x_2, t; x', 0) \psi(x', 0) dx'$$

and we want the amplitude between sources located at $x_1$ and $x_2$, so $\psi(x', 0) = \alpha \delta(x' - x_1)$ whence

$$\psi_{12} = \alpha \sqrt{\frac{m}{2\pi \hbar i t}} \exp\left[\frac{i m x_{12}^2}{2\hbar t}\right] = \alpha \sqrt{\frac{m}{2\pi \hbar i t}} \exp\left[\frac{i p x_{12}}{2\hbar}\right]$$

where $x_{12}$ is the spatial distance between sources $J_1$ and $J_2$, $t$ is the interaction time and $p = \frac{mx_{12}}{t}$. Assuming the interaction time is large compared to the 'exchange' particle's characteristic time so that $x_{12}$ is large compared to $\frac{\hbar}{p}$ we have

$$\psi = \psi_{23} + \psi_{43} \propto \exp\left[\frac{ipx_{23}}{2\hbar}\right] + \exp\left[\frac{ipx_{43}}{2\hbar}\right] \quad (9)$$

as the Schrödinger counterpart to Eq. (8), whence we infer

$$\frac{p}{2\hbar}x_{ik} = \frac{\Gamma_i d_{ik} j_k}{2\pi\hbar} \quad (10)$$

Assuming the impulse $j_k$ is proportional to the momentum transfer $p$, we have

$$x_{im} \propto \frac{\Gamma_i k_{im}}{\left(k_{im}^2 - (\omega_o^2 m - k)^2\right)} \quad (11)$$

relating the spatial separation $x_{im}$ of the trans-temporal objects $J_i$ and $J_m$ to their intrinsic ($m$, $k$, $\omega_o$) and relational ($k_{im}$) 'dynamical' characteristics.

As we stated in section 1, the metric of Eq. (11) provides spatial distance only between interacting ($k_{im} \neq 0$) trans-temporal objects, in stark contrast to the metric field of relativity theory which takes on values at each point of the differentiable spacetime manifold, even in regions where the stress-energy tensor is zero. And, as is clear from our presentation, there is no 'exchange' particle or wave (of momentum $p$ or otherwise) moving 'through space' from the source to the detector to 'cause' a detection event. Accordingly, there is no concept of spatial distance in spacetime regions where the stress-energy tensor vanishes. Thus, our simple analysis of Feynman's "mystery," in accord with Pauli's dictum concerning the articulation of composition and separability, resonates strongly with Smolin's sentiment that the nature of time and QM's foundational issues may be highly relevant to quantum gravity.

# REFERENCES

1. R.P. Feynman, R.B. Leighton and M. Sands, *The Feynman Lectures on Physics III, Quantum Mechanics* (Addison-Wesley, Reading,1965), p. 1-1.
2. W. Pauli, *Scientific Correspondence with Bohr, Einstein, Heisenberg a.o.*, *Vol 2, 1930-1939*, edited by Karl von Meyenn (Springer-Verlag, Berlin, 1985), pp. 402-404.
3. L. Smolin, *The Trouble with Physics* (Houghton Mifflin, Boston, 2006), p. 256.
4. W.M. Stuckey, M. Silberstein, and M. Cifone, Phys. Ess. (to be published); preprint arXiv: quant-ph/0503065.
5. S. French and D. Krause, *Identity in Physics: A Historical, Philosophical and Formal Analysis* (Clarendon, Oxford, 2006), p. 19.
6. D. Howard, in *Potentiality, Entanglement and Passion-at-a-Distance*, edited by R.S. Cohen *et al*. (Kluwer Academic, Great Britain, 1997), p. 122.
7. Ibid, p. 129.
8. A. Zee, *Quantum Field Theory in a Nutshell* (Princeton U.P., Princeton, 2003), p. 18.
9. T. Toffoli, *Int. J. Theor. Phys.* **42**, #2, 363-381 (2003).
10. C.W. Misner, K.S. Thorne & J.A. Wheeler, *Gravitation* (W.H. Freeman, San Francisco, 1973), p. 364.
11. W.M. Stuckey, *Phys. Ess.* **12**, 414-419, (1999).
12. M. O. Scully and K. Druhl, Phy. Rev. A **25**, 2208 (1982).
13. R. Shankar, *Principles of Quantum Mechanics*, 2[nd] Ed (Plenum Press, New York, 1994), p. 226.